\documentclass[twocolumn,english,prl,showpacs,preprintnumbers,superscriptaddress]{revtex4-2}
\usepackage[utf8]{inputenc}
\usepackage{color}
\usepackage{amsmath,amssymb}
\usepackage{amssymb}
\usepackage{graphicx}
\usepackage{textcomp}
\usepackage{dcolumn}
\usepackage{bm}
\usepackage[right]{eurosym}
\usepackage{float}
\usepackage[english]{babel}
\usepackage{blindtext}
\usepackage{babel}
\usepackage[normalem]{ulem}

\begin{document}

\title{Spectroscopically resolved resonant interatomic Coulombic decay in photoexcited large He nanodroplets}

\author{L. Ben Ltaief}

\affiliation{Department of Physics and Astronomy, Aarhus University, 8000 Aarhus C, Denmark}
\author{K. Sishodia}

\affiliation{Quantum Center of Excellence for Diamond and Emergent Materials and Department of Physics, Indian Institute of Technology Madras, Chennai 600036, India}

\author{R. Richter}
\affiliation{Elettra-Sincrotrone Trieste, 34149 Basovizza, Trieste, Italy}

\author{B. Bastian}
\affiliation{Department of Physics and Astronomy, Aarhus University, 8000 Aarhus C, Denmark}
\author{J. D. Asmussen}
\affiliation{Department of Physics and Astronomy, Aarhus University, 8000 Aarhus C, Denmark}

\author{S. Mandal}
\affiliation{Indian Institute of Science Education and Research, Pune 411008, India}

\author{N. Pal}
\affiliation{Elettra-Sincrotrone Trieste, 34149 Basovizza, Trieste, Italy}

\author{C. Medina}
\affiliation{Institute of Physics, University of Freiburg, 79104 Freiburg, Germany}
\author{S. R. Krishnan}
\affiliation{Department of Physics, Indian Institute of Technology, Madras, Chennai 600 036, India}
\author{K. von Haeften}
\affiliation{DESY, Notkestr. 85, 22607 Hamburg, Germany}
\altaffiliation{Present address: Kanano GmbH, Sedanstr. 14, 89077 Ulm, Germany}
\author{M. Mudrich}
\affiliation{Department of Physics and Astronomy, Aarhus University, 8000 Aarhus C, Denmark}
\email{ltaief@phys.au.dk and mudrich@phys.au.dk}
\begin{abstract}

Interatomic Coulombic decay (ICD) processes play a crucial role in weakly bound complexes exposed to intense or high-energy radiation. Using large helium nanodroplets, we demonstrate that ICD is efficient even when the droplets are irradiated by weak synchrotron radiation at relatively low photon energies. Below the ionization threshold, resonant excitation of multiple centers efficiently induces resonant ICD as previously observed for intense pulses [A. C. LaForge et al., PRX 11, 021011 (2021)]. More surprisingly, we observe ICD even above the ionization threshold due to recombination of photoelectrons and ions into excited states which subsequently decay by ICD. This demonstrates the importance of secondary processes, in particular electron scattering and recombination, in inducing ICD in extended condensed phase systems. In addition, we show that ICD can serve as a diagnostic tool for monitoring the relaxation dynamics of highly-excited and ionized weakly-bound nanosystems. 

\end{abstract}

\date{\today}

\maketitle

\section{Introduction}

When matter is exposed to ionizing radiation, both primary ionization and secondary processes may occur. In biological tissue, radiation damage is mostly induced by the latter, \textit{e.~g.} by multiple scattering of the primary photoelectron in the medium followed by dissociative attachment of low-energy electrons to vital biomolecules~\cite{Boudaiffa:2000}. Another process creating slow, genotoxic low-energy electrons is interatomic Coulombic decay (ICD), where the energy deposited in one atom or molecule is transferred to another which in turn is ionized~\cite{Cederbaum:1997}. 

ICD has been discovered and characterized in detail for small van-der-Waals molecules and clusters~\cite{Hergenhahn:2011}. More recently, the focus has shifted to more relevant condensed-phase systems such as liquid water~\cite{Jahnke:2020,Zhang:2022}. There, the light-matter interactions are more complex and the processes causing radiation damage are harder to decipher; in particular electron scattering tends to obscure the signatures of ICD in electron spectra~\cite{malerz2021low}.

In this work, we show that large He nanodroplets are efficiently multiply excited by weak quasi-continuous synchrotron radiation leading to the resonant variant of ICD, resonant ICD~\cite{LaForge:2021}. Moreover, we find that elastic electron scattering can facilitate ICD by inducing electron-ion recombination into highly-excited states which subsequently decay by resonant ICD. He nanodroplets are a special type of condensed phase system owing to their quantum fluid nature~\cite{Toennies:2001}; Atoms and molecules inside them are highly mobile, and their electron spectra are often well-resolved~\cite{peterka:2007,Kelbg:2019,Ltaief:2019,Ovcharenko:2020,LaForge:2021}. However, electron scattering leading to low-energy electrons and electron-ion recombination occurs in other types of condensed-phase systems as well~\cite{kirm2003prompt}. In particular, the decay of multiple excited states or excitons by ICD-type processes has been observed for solid rare-gas clusters~\cite{iablonskyi2016slow,Serdobintsev:2018}, nanoplasmas~\cite{SchuetteCorrelated:2015,oelze:2017,Kelbg:2019,Kelbg:2020}, solid nanostructures~\cite{soavi2016exciton}, and thin films~\cite{lewis2006singlet,kumar2014exciton,delport2019exciton}.

He nanodroplets have previously proven well-suited as model system for elucidating ICD and related processes. In those studies, mainly either high-energy photons were used to excite into high-lying or ionized states of He~\cite{LaForgePRL:2016,Shcherbinin:2017,wiegandt2019direct,Ltaief:2020,ltaief2023efficient}, or intense pulses were used to multiply excite or ionize the droplets~\cite{LaForge:2014,Ovcharenko:2014,SchuetteCorrelated:2015,oelze:2017,Kelbg:2019,Kelbg:2020,Ovcharenko:2020,LaForge:2021,michiels2021enhancement}. Using extreme ultraviolet (EUV) pulses from a tunable free-electron laser (FEL), the transition from the regime of ICD of weakly excited He droplets to the regime of ultrafast collective autoionization (CAI) of multiply excited He droplets was tracked~\cite{Ovcharenko:2014,Ovcharenko:2020}. EUV-pump, UV-probe studies of multiply excited He droplets indicated that ICD predominantly occurs in pairs of nearest-neighbor He$^*$ excited atoms within $\gtrsim 0.4~$ps, facilitated by the merging of void bubbles forming around each He$^*$~\cite{LaForge:2021}.

An important aspect of the present study is that only weak radiation with photon energies below or just above the ionization threshold of He is used for inducing resonant ICD. The He droplets are multiply excited or ionized owing to their large size $\gtrsim20~$nm and absorption cross section $\gtrsim 2\times 10^5~$\AA$^2$. In such bulk-like systems, inelastic and multiple elastic scattering efficiently slows down photoelectrons such that they are recaptured by the photoions to populate both fluorescing and metastable states, denoted as He$^*$~\cite{asmussen2023dopant,asmussen2023electron,ltaief2023efficient}. Resonant ICD then proceeds in the droplets according to the reaction $\mathrm{He}^* + \mathrm{He}^* \rightarrow \mathrm{He} + \mathrm{He}^+ + e_\mathrm{ICD}^-$~\cite{Kuleff:2010}. Our results indicate that this ICD reaction most likely occurs for pairs of metastable $\mathrm{He}^*$ that have migrated to the surface of the He droplets.

Additionally, by measuring the energy of the emitted ICD electron $e_\mathrm{ICD}^-$, we gain detailed insight into the relaxation of the photo-excited system. We find that different states of He$^*$ are populated prior to ICD in different regimes of resonant excitation, autoionization, or direct photoionization of the droplets. This is in agreement with previous studies of the relaxation dynamics of singly excited He droplets, which have shown that electronically excited He droplets relax into the lowest excited singlet state 1s2s\,$^1$S within $\lesssim 1~$ps~\cite{Ziemkiewicz:2015,mudrich:2020,Asmussen:2021}. When the droplets are excited above their adiabatic ionization energy $E_i^\mathrm{drop}\approx 23$~eV, additionally triplet states are populated by electron-ion recombination which relax by fluorescence emission and droplet-induced electronic relaxation into the metastable 1s2s\,$^3$S state~\cite{Haeften:1997,Asmussen:2021}. Surprisingly, the ICD spectra reveal that in large He droplets, electronic relaxation into the 1s2s\,$^1$S state occurs even well above $E_{i}^\mathrm{drop}$ and recombination into the 1s2s\,$^3$S state occurs up to several eV above the vertical ionization threshold of He, $E_i=24.6~$eV. Thus, in extended systems multiple electron scattering and electron-ion recombination is another efficient route to creating multiple excitations which efficiently decay by ICD.

\section{Experimental setup}
\begin{figure}[!h]
	\center
	\includegraphics[width=0.9\columnwidth]{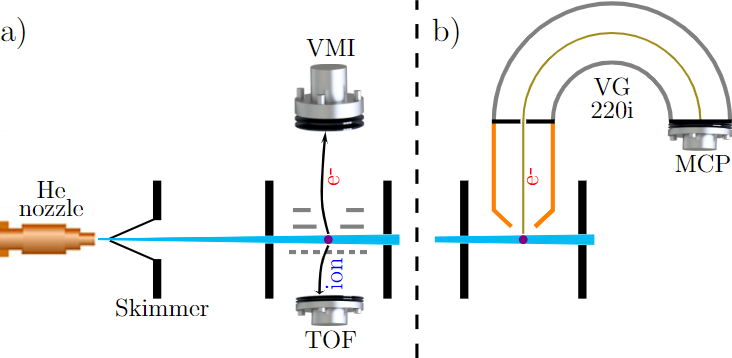}\caption{\label{Fig1} Sketch of the experimental setups used in this work. a) He nanodroplet beam source and coincidence velocity-map imaging (VMI)-time-of-flight (TOF) spectrometer. b) Hemispherical electron analyser (HEA) coupled to a microchannel plate (MCP) detector.}
\end{figure}

To probe the ICD electrons emitted from He droplets at variable photon energy, a He nanodroplet apparatus combined with a photoelectron-photoion coincidence velocity-map imaging (PEPICO-VMI) detector~\cite{OKeeffe:2011} was used at the GasPhase beamline of the Elettra synchrotron facility in Trieste, Italy. Kinetic energy distributions of electrons were inferred from the VMIs using the MEVELER inversion method~\cite{Dick:2014}. In a second arrangement, a hemispherical electron analyzer (HEA, model VG-220i) with a resolution of $<0.1$~eV was mounted at the magic angle and combined to the He nanodroplet apparatus to measure high-resolution ICD electron spectra (see Fig.~\ref{Fig1}). 

The He droplet apparatus has been described in detail elsewhere~\cite{Buchta:2013,BuchtaJCP:2013}. Briefly, a continuous beam of He nanodroplets of variable droplet radii ranging from $R= 5$~nm for droplets containing an average number of He atoms $\langle N\rangle\sim 10^4$ up to $R= 75$~nm ($\langle N\rangle\sim 10^8$) is generated by expanding He out of a cryogenic nozzle at a temperature ranging from 16 down to 8~K at 50~bar of He backing pressure. A mechanical chopper is used for discriminating the He droplet beam-correlated signals from the background.

In this study, the photon energy was tuned across the He absorption resonances and across $E_i$, \textit{i.~e.} in the range $h\nu = 21.0$ - $28.0$~eV. The use of a variable angle spherical grating monochromator ensured narrow-band radiation with a time-averaged photon flux $\Phi\approx 5\times 10^{11}~$s$^{-1}$. At photon energies $h\nu\leq 21.6$~eV, a Sn filter was inserted in the beamline to suppress higher-order radiation.

\section{Results and discussion}
\subsection{Total electron and EUV fluorescence spectra}
The strongest resonant absorption bands of large He nanodroplets are those correlating to the 1s2s\,$^1$S and 1s2p\,$^1$P states of He atoms at photon energies $h\nu=21.0$ and $h\nu=21.6$~eV, respectively~\cite{Joppien:1993,Buchta:2013}. As these excitation energies stay below the adiabatic ionization energy $E_i^\mathrm{drop}$, no direct electron emission is expected. Nevertheless, high yields of electrons are detected when the size of the He nanodroplets exceeds $R\approx 20$~nm.

\begin{figure}[!h]
	\center
	\includegraphics[width=1.0\columnwidth]{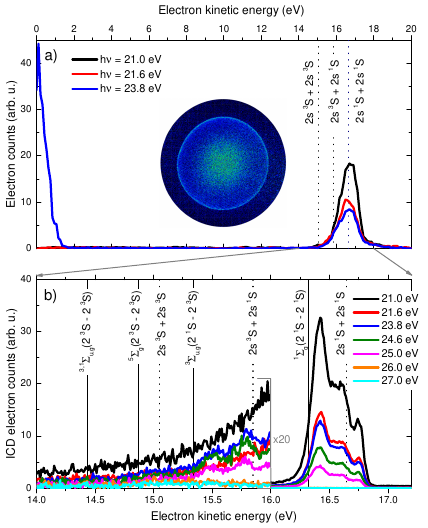}\caption{\label{Fig2} Background-subtracted total electron spectra measured for pure large He nanodroplets ($R= 75$~nm) at different photon energies near the ionization energy of He using the VMI spectrometer a) and the HEA b). The inset in a) shows a raw total VMI recorded at $h\nu=21.0$~eV. For better visibility, all of the spectra shown in b) in the kinetic energy range 14.0 - 16.0~eV are scaled up by a factor of 20. All electron spectra are normalized to the flux of the incident EUV photon beam.}
\end{figure}

Electron VMIs measured under these conditions display a sharp-edged, perfectly isotropic ring structure, see the inset in Fig.~\ref{Fig2}~a) recorded at $h\nu=21.0$~eV. Fig.~\ref{Fig2}~a) shows the total electron spectra inferred from the VMIs recorded at $h\nu=21.0$~eV (black line), $h\nu=21.6$~eV (red line) and $h\nu=23.8$~eV (blue line). All electron spectra exhibit a sharp peak around 16.6 eV. At $h\nu$= 23.8~eV, an additional narrow feature close to zero electron kinetic energy is present in the spectrum which is due to autoionization of superexcited He droplets~\cite{Froechtenicht:1996, Peterka:2003,peterka:2007}. The peak present in all three spectra is centered at the electron energy expected for ICD of two He$^*$ in 1s2s\,$^1$S states, $E_
e = 2 E_{1S}-E_i = 2\times 20.6~\mathrm{eV}-24.6~\mathrm{eV}=16.6~\mathrm{eV}$, irrespective of the photon energy. Here, $E_{1S}$ is the excitation energy of the 1s2s\,$^{1}$S state. This indicates that the He droplets mostly relax from the initially excited 1s2s, 1s2p and 1s3p-correlated states of the droplet into the lowest excited 1s2s\,$^1$S singlet state of the He$^*$ atom prior to ICD. Fast vibronic relaxation preceding ICD has been observed before~\cite{Ltaief:2019,LaForge:2021}. This sets an upper bound to the ICD decay time to $\gtrsim 1~$ps, the relaxation time of electronically excited He droplets previously measured by time-resolved photoelectron spectroscopy~\cite{Ziemkiewicz:2015,mudrich:2020,Asmussen:2021}. 

This ICD process was previously observed using FEL pulses at much higher intensity $\gtrsim10^{10}$~Wcm$^{-2}$~\cite{Ovcharenko:2020,LaForge:2021}, but not unambiguously using synchrotron radiation~\cite{Peterka:2003,peterka:2007,Wang:2008,BuchtaJCP:2013,Shcherbinin:2018,LaForge:2019,Ltaief:2019,Mandal:2020}. In the present experiment the He droplets were produced by supercritical expansion of liquid He; in this regime the droplets are much larger (radius $R > 10~$nm, $\langle N\rangle > 10^5$ He atoms per droplet) than those conventionally used for He-nanodroplet isolation spectroscopy, $\langle N\rangle\lesssim 10^4$~\cite{Toennies:2004}. Accordingly, their absorption cross section is larger and the rate of resonant absorption of a droplet with \textit{e.~g.} $\langle N\rangle= 10^6$ is $r_\mathrm{abs}=\sigma_{2p}\langle N\rangle\Phi/w^2\approx 8\times 10^3$~s$^{-1}$. Here, the photon beam radius is $w\approx 400~\mu$m and the absorption cross section at the 1s2p-resonance of He droplets at $h\nu=21.6$~eV is estimated to $\sigma_{2p}=25$~Mbarn~\cite{BuchtaJCP:2013,Ovcharenko:2014}.  Accordingly, the probability that this droplet resonantly absorbs one photon during its flight through the interaction region is $p_1 = \sigma_{2p}\langle N\rangle t_\mathrm{tr} \Phi/w^2 \approx 1$\,\% for a transit time of the droplets through the focus $t_\mathrm{tr}\approx 1~\mu$s. As it takes two photons to excite a pair of He$^*$ atoms in one He droplets which decay by ICD, the ICD rate is $r_\mathrm{ICD}=r_\mathrm{abs}/2$~\cite{Kuleff:2010}. For an estimated number density of He droplets in the jet of $n_\mathrm{HeN}\sim 10^6~$cm$^{-3}$ and an active length of the focal volume of $d\approx 2$~mm, the total ICD rate is $R_\mathrm{ICD}=r_\mathrm{ICD} n_\mathrm{HeN}w^2 d\approx 10^6$~s$^{-1}$. This value roughly matches the rate of detected ICD electrons in the experiment considering the detection efficiency of the HEA is $\sim 10^{-3}$.

Note that $R_\mathrm{ICD}\propto\Phi$ scales linearly with photon flux $\Phi$ although two or more photons have to be absorbed by one He droplet to induce ICD. Only for much higher intensity as in previous FEL experiments would the excited-state population and hence $R_\mathrm{ICD}$ be saturated~\cite{LaForge:2014,Ovcharenko:2014}. This linear dependency of $R_\mathrm{ICD}$ on $\Phi$ is experimentally confirmed, see Fig.~1 in the supplemental material (SM). When varying the photon flux by gradually opening and closing the exit slit of the monochromator and measuring all ICD electrons produced at $h\nu = 21$~eV, we observe essentially a linear dependency over more than one order of magnitude variation of the photon flux, irrespective of the He nanodroplets size.

Using the HEA, the ICD feature seen at the electron energy $E_e=16.6$~eV in the VMI spectra are much better resolved. Fig.~\ref{Fig2}~b) shows high-resolution total electron spectra measured using the HEA at various photon energies below and above $E_i$. The spectra clearly show a substructure of the ICD peak. Additionally, a low-amplitude wing structure extends from $E_e=16.2$~eV down to about 14~eV, indicating that 1s2s\,$^3$S-excited He$^*$ atoms and correlated states of the He$_2^{**}$ dimer contribute to the ICD signal to a small extent. At $h\nu$= 21.0 and 21.6~eV, the electron spectra exhibit one peak at $E_e=16.3$~eV originating from ICD of pairs of He$^*$ in the $^1\Sigma_g$ state correlating to two atoms each in the 1s2s\,$^1$S singlet state~\cite{LaForge:2021}, and a shoulder that extends up to 16.8~eV featuring two smaller peaks. Electron spectra recorded at 23.8~eV $\le$ $h\nu$ $\le$ 24.6~eV exhibit three small maximima at $E_e=15.0$, $15.5$ and $15.8$~eV in addition to the main ICD features. The weak peak at 15.0 ~eV is assigned to $^5\Sigma_g$/$^3\Sigma_{u,\,g}$ states correlating to two  He$^*$ atoms in 1s2s\,$^3$S triplet states~\cite{LaForge:2021}. The peaks at 15.5~eV and 15.8~eV can be ascribed to ICD out of $^3\Sigma_{u,\,g}$ states correlating to a mixed pair of metastable He$^*$($^1$S, $^3$S). The structure of these two peaks resembles very well the structure of electron kinetic energy features observed before in earlier low-energy binary He$^*$($^1$S, $^3$S) collisions experiment \cite{muller:1987, Mueller:1991}. Interestingly, all of these three small peaks that involve the 1s2s\,$^3$S states only appear in the electron spectra at $h\nu \geq E_i^\mathrm{drop}$. The main ICD features around $E_e$ = $16.5~$eV are still visible up to $h\nu$ = $25.0$~eV and disappear for $h\nu$ $\geq$ $26.0$~eV whereas the small peaks at $E_e<16$~eV are still faintly visible.

This indicates that autoionization or photoionization followed by electron-ion recombination preferentially populate the metastable 1s2s$^3$S state which in turn undergoes ICD by interaction with another 1s2s$^3$S or 1s2s$^1$S He atom~\cite{Haeften:1997,Asmussen:2021}.
The structure of the high-resolution ICD spectra at $E_e = 16.0$ - 17.0~eV in Fig.~2~b) will be discussed in more detail in Sec.~III C.

\begin{figure}[!h]
	\center
	\includegraphics[width=0.9\columnwidth]{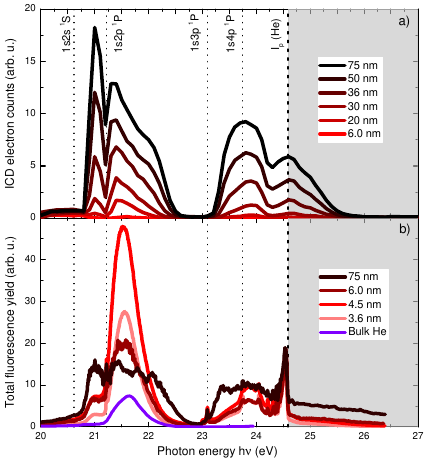}\caption{\label{Fig3} Total yield spectra of ICD electrons a) and EUV fluorescence b) measured for He nanodroplets of various sizes. The purple line is the absorption spectrum of bulk liquid He taken from~\cite{Surko:1969}. The EUV fluorescence data shown in b) are reproduced from~\cite{vonHaeften:2011}. 
 }
\end{figure}
To get an overview of the ICD efficiency across the entire photoexcitation spectrum of He droplets, the photon energy was tuned from $h\nu = 20$ to 26.5~eV while measuring the yield of all ICD electrons with the HEA. Fig.~\ref{Fig3}~a) shows the HEA signal integrated in the electron energy range $E_e=16$ - 17~eV for He droplets of various sizes in the range $R = 6$ - $75$~nm.

The ICD yield exhibits four main features: A sharp peak at $h\nu=21.0$~eV associated with the 1s2s\,$^1$S droplet resonance and two broad features peaked at $h\nu=21.4$~eV and $h\nu=23.8$~eV associated with the 1s2p\,$^1$P and 1s3p/1s4p droplet states, respectively~\cite{Joppien:1993,vonHaeften:2011}. A fourth maximum appears at $h\nu =E_i=24.6~$eV, where a high density of Rydberg states is expected. These features are invisible for small droplets ($R < 20$~nm) [see red line in Fig.~\ref{Fig3}~a)]. They are clearly visible at $R = 20$~nm and become more and more pronounced when the droplet radius is further increased to $R=75$~nm [black line in Fig.~\ref{Fig3}~a)]. 

For comparison, Fig.~\ref{Fig3}~b) shows previously measured EUV fluorescence yield spectra of small He droplets (red lines)~\cite{vonHaeften:2011}. The dark line, showing the EUV fluorescence spectra for large He droplets ($R = 75$~nm), features a similarly distorted resonance structure as the ICD spectra; the 1s2s$^1$S-correlated peak at $h\nu=21.0$~eV is enhanced, the 1s2p\,$^1$P-correlated peak at $h\nu=21.6$~eV is asymmetrically broadened, and the 1s3p/1s4p-correlated feature around $h\nu=23.8$~eV is enhanced as compared to the fluorescence spectra for small He droplets ($R\le 6$~nm). The feature around $h\nu =24.6~$eV remains sharp in the fluorescence spectra for all sizes, likely due to the contribution of free He atoms and small He clusters accompanying the He droplets in the jet. Note that the fluorescence spectra contain contributions from all singly excited He species decaying to the ground state, whereas ICD spectra are selective to large He droplets which absorb at least two photons in the course of their interaction with the photon beam.

It is also interesting to note that, contrary to the ICD yields shown in Fig.~\ref{Fig3}~a), the EUV fluorescence yields for large He droplets ($R>4.5$~nm) are reduced in intensity. This suggests that for large He droplets where two or more absorption events per droplet become probable, a large fraction of the excited He atoms decay by ICD instead of decaying by fluorescence emission. Hence, ICD competes with fluorescence emission; it would be interesting to quantify the branching ratio of ICD and fluorescence emission. In future experiments both channels should be measured simultaneously.

\begin{figure}[!h]
	\center
	\includegraphics[width=1.0\columnwidth]{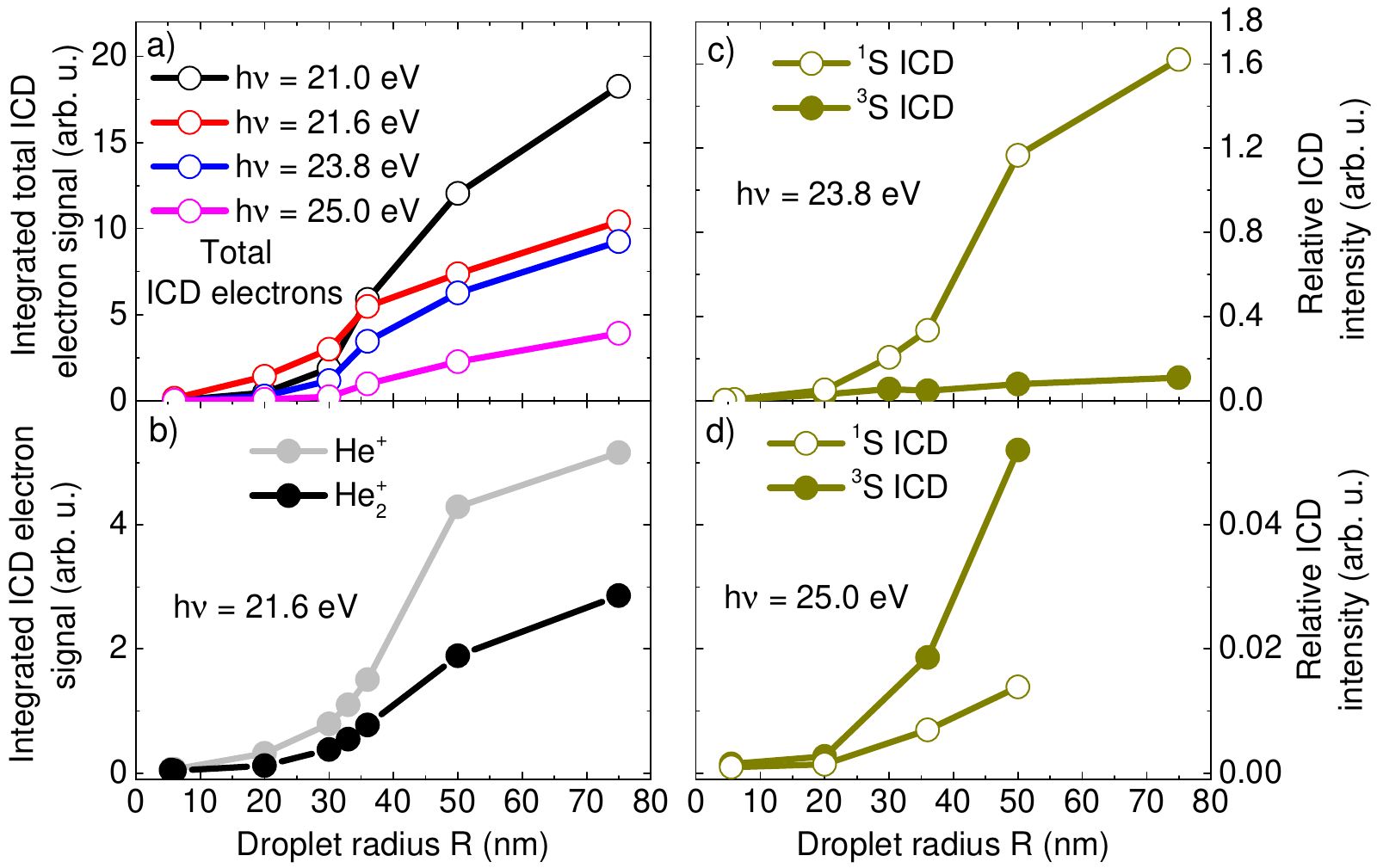}\caption{\label{Fig4} a) Droplet size-dependent total ICD electron yields measured at various photon energies. b) Integrated ICD signal measured in coincidence with He$^+$ (gray curve) and He$_2^+$ (black) at $h\nu =21.6$~eV and as a function of droplet size. Droplet size-dependent relative intensity of $^1$S ICD \textit{vs.} $^3$S ICD measured at $h\nu =23.8$~eV c) and at $h\nu = 25.0$~eV d).}
\end{figure}

The most striking feature of the ICD spectra of large He droplets is the enhanced intensity of the peak at $h\nu=21.0$~eV which becomes the highest peak for droplet sizes $R>36~$nm. Note that the absorption cross section at the 1s2s\,$^1$S resonance of medium-sized He droplets is smaller than the absorption cross section of the 1s2p\,$^1$P resonance by a factor $\approx 1/7$~\cite{Joppien:1993}. The enhancement of the peak at $h\nu=21.0$~eV for large droplets can also be seen from the total electron spectra shown in Fig.~\ref{Fig2}~b) and in Fig.~\ref{Fig4}~a) which compares droplet size-dependent ICD electron yields measured at $h\nu=21.0$~eV to those measured at higher photon energies.

The change of the structure of the ICD and EUV fluorescence spectra for increasing He droplet sizes $R>20~$nm may lead to the assumption that the spectra for large He droplets approach the characteristic absorption spectrum of bulk superfluid He. Intriguingly, the latter (measured in reflection from the surface of liquid He) resembles the spectra measured for small He droplets, though, see the purple line in Fig.~\ref{Fig3} b)~\cite{Surko:1969}. This indicates that the modified peak structure for large He droplets is related to their intrinsic properties. In particular, nano-optical effects, as observed in other types of nanoparticles~\cite{signorell2016nanofocusing}, may be expected to influence the absorption and emission spectra in this range of photon energy and He droplet size; around the 1s2p\,$^1$P resonance, the index of refraction is expected to significantly deviate from 1 by about $\pm 0.5$~\cite{rupp2017coherent} which facilitates nano-focusing effects. Additionally, the wavelength of the EUV radiation $\lambda=59~$nm matches the He droplet size studied here which may lead to resonance effects and the enhancement of the light absorption. Further experiments and simulations should be done to investigate this interesting nano-optical system in detail.

At $h\nu>E_i$, where direct emission of photoelectrons from He droplets is observed~\cite{peterka:2007,BuchtaJCP:2013}, one would probably not expect to detect any ICD and EUV fluorescence signals. However, both ICD and EUV fluorescence are detected up to $h\nu = 26$~eV or even higher for large He droplets with $R\gtrsim20~$nm. The ICD electron yield appears as a broad feature peaked at $h\nu = E_i = 24.6$~eV which reaches up to $h\nu = 26$~eV. The EUV fluorescence signal appears as a tail that continuously drops even beyond $h\nu =26$~eV. This indicates that electron-ion recombination is effective for photon energies exceeding $E_i$ by several eV~\cite{Haeften:1997,Carata:1999,Coman:1999,Peterka:2003,Urbain:2004,Haeften:2005,Pedersen:2005,Royal:2007,Buhr:2008,Asmussen:2021}. Electron-ion recombination may be expected to be particularly efficient for $h\nu\lesssim E_i + V_0 = 25.6$~eV where $V_0\approx 1~\mathrm{eV}$ is the gap to the conduction band edge of superfluid He~\cite{mauracher2018cold,Haeften:2005,Buchenau:1991}. Photoelectrons created with kinetic energy $E_e\lesssim V_0$ promptly localize in the droplet by forming bubbles, which facilitates the recombination with their parent ions. The excited He$^*$ atoms formed in this way subsequently decay either by fluorescence emission or by ICD.

\subsection{Electron-ion coincidence spectra}
\begin{figure}[!h]
	\center
	\includegraphics[width=0.8\columnwidth]{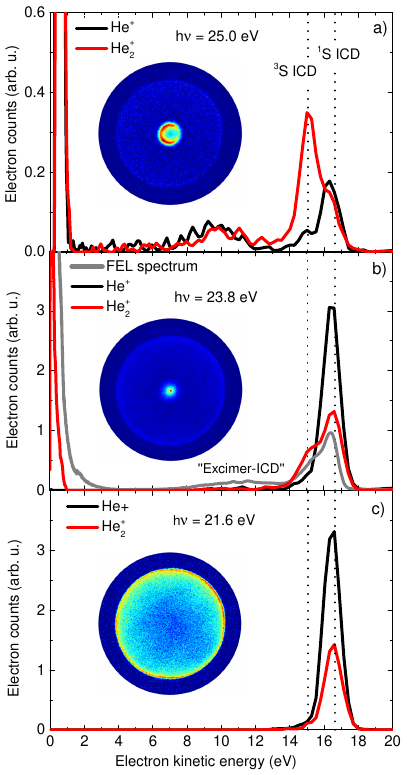}\caption{\label{Fig5} Background-subtracted electron spectra of large pure He nanodroplet ($R = 50$~nm) measured in coincidence with He$^+$ and He$_2^+$ ions at photon energies below and about the ionization energy of He. The two dotted lines indicate the kinetic energies of electrons expected for ICD of two He atoms in their metastable 1s2s\,$^1$S and 1s2s\,$^3$S states. Insets show the raw electron VMI's in coincidence with He$_2^+$. The polarization of the EUV light was vertical and the EUV beam was incident on the droplets from the left-hand side. The electron spectra are normalized to the flux of the incident EUV photon beam.}
\end{figure}

More detailed insights into the relaxation of large He nanodroplets are obtained from electron and ion spectra recorded by coincidence detection. Fig.~\ref{Fig5} shows typical electron spectra measured in coincidence with He$^+$ (black lines) and He$_2^+$ (red lines) for large He nanodroplets ($R=50$~nm) at $h\nu=25.0$~eV a), 23.8~eV b) and 21.6~eV c). The corresponding raw electron VMIs in coincidence with He$_2^+$ are displayed as insets. At $h\nu = 21.6$~eV, one sharp-edged, perfectly isotropic ring structure is seen due to ICD electrons. At $h\nu =23.8$~eV, the ICD ring is still present but an additional central bright spot appears which is due to emission of electrons by autoionization. At $h\nu =25$~eV, an anisotropic small ring at the center of the image is due to direct emission of photoelectrons. Note that this ring features a forward/backward asymmetry with respect to the propagation direction of the photon beam; this ``shadowing effect''~\cite{signorell2016nanofocusing} occurring  in large He nanodroplets is discussed in detail in Ref.~\cite{asmussen2023electron}. 

At $h\nu$= 21.6~eV, both electron spectra measured in coincidence with He$^+$ and He$_2^+$ exhibit only one main peak due to ICD of two He$^*$ atoms in the $^1$S state which is populated by electronic relaxation after exciting the He droplets to the 1s2p\,$^1$P resonance~\cite{Joppien:1993,Buchta:2013}. At $h\nu =23.8$~eV, the ICD electron spectra measured in coincidence with He$_2^+$ [Fig.~\ref{Fig5}~b)] feature a double-peak structure. At this photon energy, the He droplets are excited into the 1s3p and 1s4p absorption bands~\cite{Ziemkiewicz:2015,Asmussen:2021} which then decay by ultrafast electronic relaxation into the 1s2s\,$^1$S atomic state and by autoionization. The latter pathway can further lead to a dissociative electron-ion recombination, thereby preferentially populating the 1s2s\,$^3$S state~\cite{Haeften:1997,Pedersen:2005,Asmussen:2021}. Therefore the shoulder structure at about 15~eV appears at the electron energy expected for two He$^*$ in the $^3$S state decaying by ICD, $E_e = 2 E(^3S)-E_i = 2\times 19.8~\mathrm{eV}-24.6~\mathrm{eV} = 15.0~\mathrm{eV}$. The peak at near-zero electron energy is due to electrons emitted by droplet autoionization that do not recombine~\cite{Peterka:2003,peterka:2007}. 

Note that the $^3$S ICD feature is only seen in the electron spectrum measured in coincidence with He$_2^+$ and not in the electron spectrum measured in coincidence with He$^+$. In contrast, the $^1$S ICD feature is present in both coincidence electron spectra. Thus $^3$S ICD generates only He$_2^+$ ions, whereas $^1$S ICD generates both He$^+$ and He$_2^+$ ions. This indicates qualitatively that there are different scenarios for ICD in the He nanodroplets:
i) He$^*$ atoms excited in the $^1$S state are formed by electronic relaxation accompanied by the migration of the He$^*$'s to the droplet surface. There, two He$^*$ excited atoms undergo ICD with only little influence by the He droplet. Accordingly, mostly He$^+$ ions are produced. Low yields of He$_2^+$ are likely due to the binding of a He atom to the He$^+$ product as it escapes from the droplet. ii) He$^*$ atoms excited in the $^3$S state are formed as a result of dissociation of excited He$_2^*$'s being produced by electron-He$_2^+$ ion recombination that mainly occurs in the bulk of the droplets. Two He$^*$ excited atoms formed in this way undergo ICD prior to their ejection to the droplet surface. Therefore, the resulting He$^+$ product has a high chance of picking up another He atom to form a He$_2^+$ which is eventually ejected from the droplet by a non-thermal process. Associative ionization, \textit{i.~e.} the formation of stable He$_2^+$ by ICD can be ruled out as it is a minor channel \cite{muller:1987}.

When the photon energy is tuned across $E_i$ up to $h\nu = 25$~eV, direct photoemission becomes the dominant process, see the sharp peak at $E_e =0.4$~eV in Fig.~\ref{Fig5}~a) which matches the expected position of the photoline at $E_e =h\nu - E_i$. Remarkably, ICD features are still clearly visible implying the presence of two or more neutral excitations in one droplet. In this regime, ICD out of the 1s2s\,$^3$S state is the main indirect decay channel in the electron spectra measured in coincidence with He$_2^+$. Thus, electron-ion recombination which populates the 1s2s\,$^3$S state appears to contribute more abundantly to the electron-He$_2^+$ ion coincidences than electronic relaxation which mainly leads to the 1s2s\,$^1$S state. This can also be seen from the He-droplet size dependence of the $^1$S and $^3$S ICD components measured at $h\nu = 25.0$~eV, see Fig.~\ref{Fig4}~d). Beyond the onset of ICD at a droplet radius $R\approx 20$~nm, ICD of the $^3$S state clearly dominates over $^1$S ICD. The opposite is true at the photon energy $h\nu = 23.8$~eV just above $E_i^\mathrm{drop}$, see Fig.~\ref{Fig4}~c) which is based on the electron spectra shown in SM Fig.~3.

\begin{figure}[!h]
	\center
	\includegraphics[width=1.0\columnwidth]{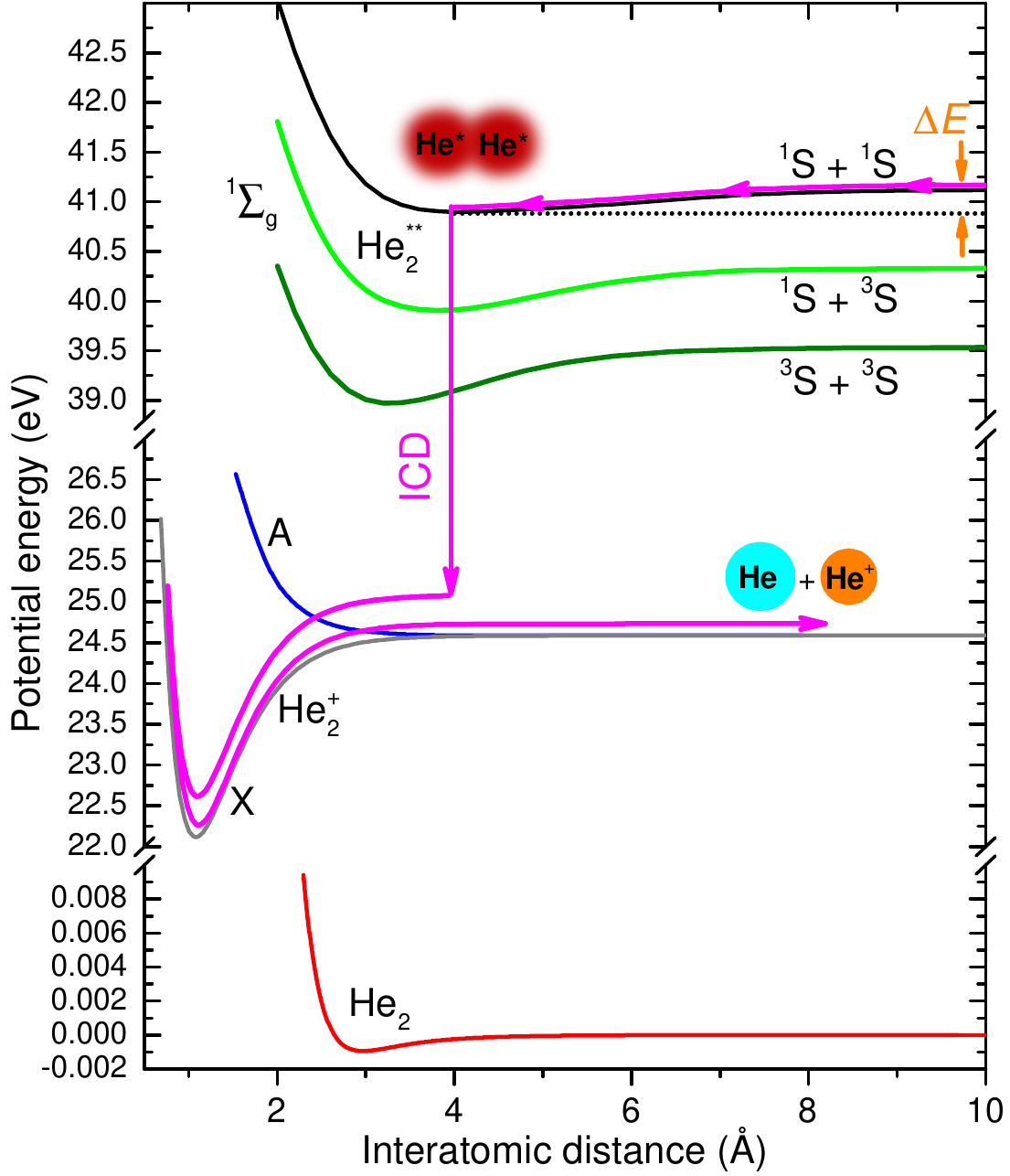}\caption{\label{Fig6} Potential energy curves of ground state He$_2$ (red line) \cite{Sheng:2020}, ground and first excited state of He$_2^+$ (grey and blue lines) \cite{Carrington:1995}, and $^1\Sigma_g$ states of doubly excited He$_2^{**}$ (black and green lines)~\cite{LaForge:2021}. The pink arrows indicate the atomic motion in the course of ICD of a pair of He$^*$'s along the He$_2^{**}$ and He$_2^+$ potential energy curves. The two red circles indicate the He$_2^{**}$ excited dimer, while the cyan and orange circles represent the ICD products -- a neutral He and a He$^+$ ion and electron, respectively. The well depth of the He$_2^{**}$ potential $\Delta E$ is the maximum potential energy converted into kinetic energy of the two atoms decaying by ICD.}
\end{figure}

This enhanced efficiency of $^3$S ICD over $^1$S ICD when detecting electron-ion coincidences can be rationalized by the atomic motion occurring in the encounter of two He$^{*}$ excited atoms along the He$^{**}_2$ potential energy curves, see Fig.~\ref{Fig6} ~\cite{LaForge:2021}. The atomic motion during the ICD process is indicated by pink arrows; see Sec.~\ref{sec:ions} for a more detailed discussion. As the potential well depths of the He($^3$S)-He($^3$S) dimer states are twice as deep as for the He($^1$S)-He($^1$S) dimer state, the ion produced by $^3$S ICD is released with higher kinetic energy and therefore it is ejected out of the He droplet more efficiently. 

The He$_2^+$ and He$^+$-correlated electron spectra recorded at $h\nu = 23.8$~eV and $h\nu = 25.0$~eV contain another weak component in the range 7 - 13~eV which was also seen in FEL experiments at $h\nu = 23.8$~eV [grey line in Fig.~\ref{Fig5}~b)]~\cite{LaForge:2021}. This feature is interpreted as due to ICD involving He$_2^*$ excimers. Interestingly, it is only observed in the electron spectra recorded at $h\nu \geq E_i^\mathrm{drop}$ and not in the electron spectra measured following photoexcitation of the He droplets into the 1s2p-correlated band at $h\nu = 21.6$~eV, see Fig.~\ref{Fig5}~c). This further entails the evidence of electron-ion recombination in leading to the formation of He$_2^*$'s upon autoionization or direct ionization of the He droplets~\cite{Buchenau:1991, ltaief2023efficient}. Note that in bulk liquid He, He$_2^*$ excimers can also be produced following primary ionization events~\cite{benderskii1999photodynamics,gao:2015,dennis:1969,hill:1971}. Both atomic and molecular triplet emission lines upon relaxation of He$_2^*$ were previously observed in fluorescence spectra of He clusters at $23.1$~eV $\leq h\nu\leq 24.6$~eV~\cite{Haeften:1997}. While electron recombination with He$_2^+$ in the droplet can be a dissociative as mentioned above, inside a He droplet fraction of He$_2^*$'s formed by recombination can also be stabilized by the cold He environment. When the stabilized He$_2^*$ excimer subsequently decays into the electronic groundstate by ICD, the amount of energy transferred to the reaction partner (He$^*$ or He$_2^*$) is significantly lower compared to ICD where an excited He$^*$ decays to the groundstate; the He$_2^*$ excitation energy is lower by 2.5~eV and the He$_2$ groundstate potential is strongly repulsive at the He$_2^*$ equilibrium distance (1.08~\AA), see the red line in Fig.~\ref{Fig6}~\cite{Fiedler:2014,Sheng:2020}. This He$_2^*$ ICD feature is also observed in large droplets at $h\nu\geq44.4~eV$ where inelastic scattering facilitates the population of excited states~\cite{ltaief2023efficient}. 

\begin{figure}[!h]
	\center
	\includegraphics[width=1\columnwidth]{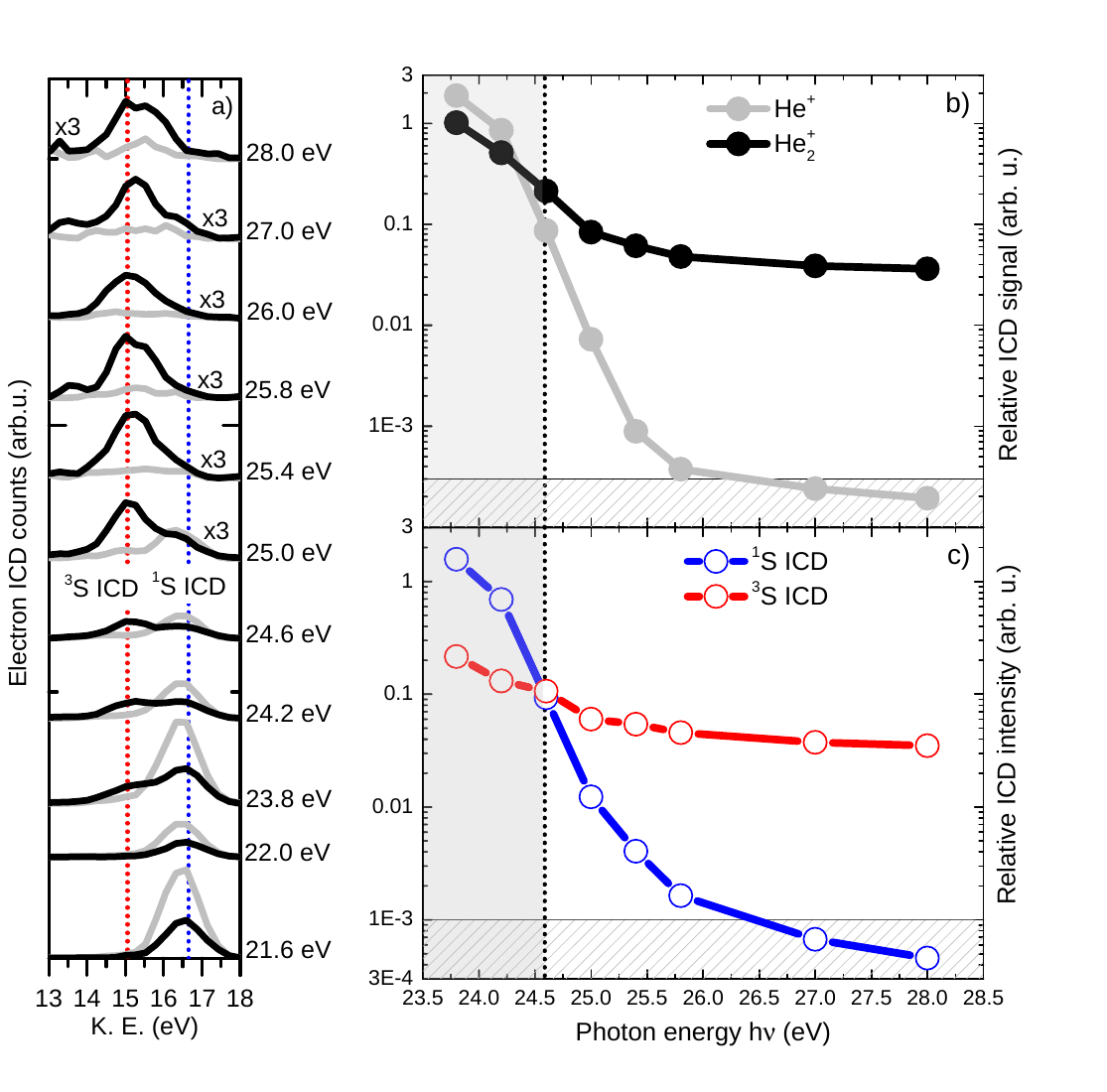}\caption{\label{Fig7} a) Background-subtracted ICD electron spectra measured in coincidence with He$^+$ (gray lines) and He$_2^+$ (black lines) for pure large He nanodroplets ($R = 50$~nm) at different photon energies below and above the He ionization energy $E_i$. All ICD electron spectra are normalized to the photon flux. The red and blue dotted lines indicate the kinetic energies of an electron emitted by ICD of the $^{3}$S and $^{1}$S states, respectively. b) Integrated yields of ICD electrons detected in coincidence with He$^+$ (gray dots) and He$_2^+$ (black dots) as a function of photon energy. (c) Ratio of $^1$S (blue dots) and $^3$S (red dots) ICD electrons to photoelectrons (near-zero energy electrons for the case of $h\nu\leq E_i$).}
\end{figure}

For even higher photon energies we expect that electron-ion recombination becomes the only way of inducing ICD, as the electron promoted into the conduction band detaches from the He$^+$ core and has to undergo multiple elastic collisions in the droplet to lose enough energy and return to the He$^+$. Indeed, when tuning the photon energy from $h\nu = 21.6$~eV to 28.0~eV, we see a transition from $^1$S ICD to $^3$S ICD, and for $h\nu>25$~eV, $^3$S ICD clearly dominates the electron-ion spectra measured in coincidence with He$_2^+$, see Fig.~\ref{Fig7}~a) and b). The ratios of $^1$S and $^3$S ICD peak areas versus photoelectrons are shown in Fig.~\ref{Fig7}~c). They are obtained from fitting Gaussian functions to the $^1$S and $^3$S ICD peaks in the electron spectra measured in coincidence with He$^+$ and He$_2^+$, and as the weighted average of their $^1$S and $^3$S ICD peak heights, respectively [see SM Sec. III]. Below the He ionization threshold, $h\nu\leq E_i$, the near-zero kinetic energy peak resulting from droplet autoionization is fitted instead of the photoline at $h\nu > E_i$. In this representation of the data, the transition from relaxation-dominated ICD, leading mostly to the $^1$S ICD peak, to recombination-dominated ICD, leading to the $^3$S peak, occurs right at $h\nu = E_i$. The $^1$S ICD signal drops to the noise level for $h\nu \gtrsim 25.5$~eV whereas recombination-induced ICD remains visible even at $h\nu = 28$~eV.

\subsection{Ion kinetic energy distributions}
\label{sec:ions}
\begin{figure}[!h]
	\center
	\includegraphics[width=1.0\columnwidth]{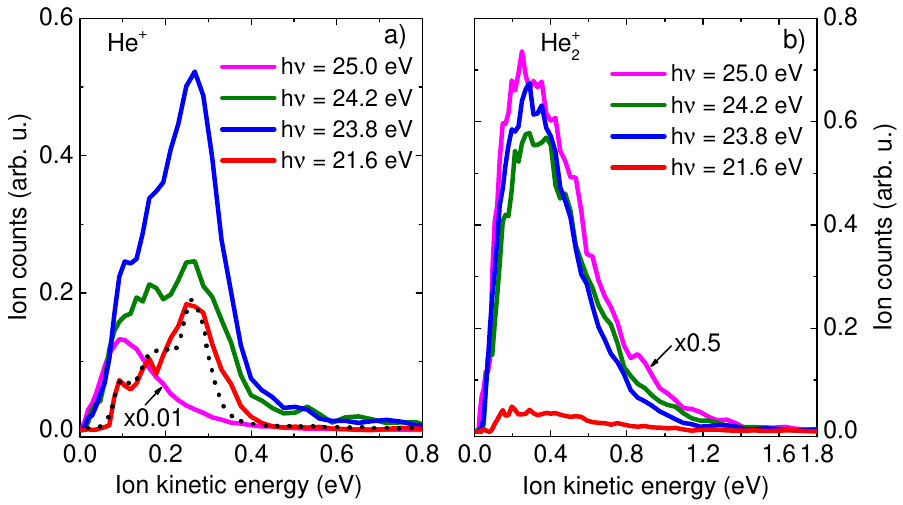}\caption{He$^+$ and He$_2^+$ ion kinetic energy distributions measured for pure large He nanodroplets of radius $R= 50$~nm at different photon energies below and above the He ionization energy. All ion spectra in a) and b) are background subtracted and normalized to the EUV photon flux. The black dotted curve in a) is obtained by linear transformation of the electron spectrum measured at $h\nu = 21.6$~eV [see red line in Fig.~\ref{Fig1} b)] according to Eq.~(\ref{eq:trafo}). \label{Fig8}}
\end{figure}

Complementary information about the relaxation dynamics of EUV-irradiated large He droplets is obtained from ion kinetic energy distributions.
Fig.~\ref{Fig8} shows kinetic energy distributions of He$^+$ and He$_2^+$ ions measured in coincidence with one electron at various photon energies below $E_i$. The He$_2^+$ ion spectra shown in Fig.~\ref{Fig8}~b) have the same structure with a broad kinetic energy distribution centered around $E_\mathrm{ion}=0.3$~eV that extends up to 1.4~eV.

The majority of He$_2^+$ ions detected at $h\nu= 23.8$, 24.0, 24.2~eV are created by autoionization (\textit{cf.} Fig.~\ref{Fig5}), and their kinetic energy distributions follow the one measured at $h\nu = 25.0$~eV or higher, see SM Fig.~2~b), in agreement with previous photoionization measurements~\cite{Shcherbinin:2019}. In contrast, ions detected at $h\nu =21.6$~eV are mainly created by ICD [\textit{cf.} Fig.~\ref{Fig5}~c)]. In both cases, (autoionization and ICD) the He$_2^+$ ions are ejected out of the He droplet by a non-thermal, impulsive process in the course of vibrational relaxation~\cite{callicoatt1998fragmentation}. 

The kinetic energy distributions of He$^+$ ions are shown in Fig.~\ref{Fig8}~a). Interestingly, all He$^+$ ion energy distributions recorded at $h\nu<E_i$ are similar in shape with a pronounced maximum at 0.27~eV. They can only originate from ionization of excited He$^*$'s by ICD following two-photon absorption by the droplets. Furthermore, the shape of these spectra clearly differs from the shape of the He$^+$ spectrum measured at $h\nu= 25.0$~eV, where only one main broad feature peaking around 0.1~eV with a tail extending to 0.5~eV is observed. The latter broad feature is also visible in the ion spectra recorded for small He droplets. Remarkably, it does not change structure when varying the He droplet size and when tuning the photon energy above $E_i$, see SM Fig.~2~a). We interpret this generic distribution of ion energies by photoionization into the repulsive $A$ state of the He$_2^+$ molecular ion, see the blue line in Fig.~\ref{Fig6}. 

For small He droplets ($R< 20$~nm), an additional sharp peak near 0~eV is present in the ion spectra recorded at $h\nu > E_i$, see SM Fig.~2~a). It is due to photoionization of the free He atoms accompanying the He droplets in the jet, as discussed in~\cite{Shcherbinin:2019}. Note that the He$^+$ and He$_2^+$ ion spectra recorded at $h\nu>E_i$ for large He nanodroplets ($R > 20$~nm) should include a contribution of He$^+$ and He$_2^+$ ions created by ICD. However, these ICD ions are hard to identify due to the overwhelming contribution from ions created by direct photoionization.

The proposed relaxation pathway of two He$^*$'s formed by absorption of two EUV photons by one He nanodroplet (resonant excitation or electron-ion recombination) leading to the ejection of an ICD electron and ion is illustrated in Fig.~\ref{Fig8}~\cite{LaForge:2021, Carrington:1995, Sheng:2020}. Following photoexcitation, the two He$^*$'s are accelerated toward each other from large interatomic distance $R$ along the attractive $^1\Sigma_g$ potential curve of the doubly excited He dimer, He$_2^{**}$; when they reach shorter distances $R$, the ICD probability rises and ICD likely occurs near the well of the potential around $R=4~$\AA, leading to the emission of an ICD electron with an energy corresponding to the difference potential between the initial He$^{**}_2$ state and the final He$_2^+$ state at the distance $R$. The maximum kinetic energy acquired by the two He$^*$ atoms is given by the depth of the potential well with respect to the He$^*$+He$^*$ atomic asymptote, $\Delta E$. As the kinetic energy acquired by the two colliding He atoms in the $^1\Sigma_g$ state is not significantly affected by the ICD process, the He and He$^+$ atoms in the final state continue their trajectory toward short $R$ where they are reflected at the hard-core potential of the He$_2^+$ ground state $X$. In this process, the ICD electron energy is reduced at the extent at which the kinetic energy of the products increases in the course of the collision; therefore the ICD electron spectrum can be transformed into a kinetic energy distribution of the He$^+$ ICD ion according to 
\begin{equation}
E_\mathrm{ion}=(2E_\mathrm{He^*}-E_i-E_e+ dE)/2.    
\label{eq:trafo}
\end{equation}
Here, $E_\mathrm{He^*}=20.62$~eV is the excitation energy of each He$^*$ and $dE$=0.3~eV is a droplet-induced energy shift. The factor $1/2$ accounts for equal sharing of the kinetic energy released to the two dissociating He atoms. This calculated ion kinetic energy distribution from the high resolution electron spectrum measured at $h\nu= 21.6$~eV [red line in Fig.~\ref{Fig2}~b)] matches the corresponding He$^+$ ion energy distribution surprisingly well, see the dotted black line in Fig.~\ref{Fig8}~a).
This indicates that the He$^+$ created by ICD are indeed ejected from the He nanodroplet by a binary collision-like process where the He and He$^+$ products dissociate without undergoing further scattering. This confirms our conjecture that ICD happens predominantly out of relaxed He$^*$ atoms that have migrated to the He droplet surface. The asymmetric three-peak structure seen in the ion spectra measured at $h\nu$ $\le$ $E_i$ is likely due to quantum interference effects in the entrance channel of colliding pair of metastable atoms as observed earlier in low-energy binary collisions~\cite{muller:1987,Mueller:1991}.

\section{Conclusion}
In summary, we have studied in detail the decay of multiply excited He nanodroplets by resonant ICD. Owing to the large absorption cross section of He droplets of sizes $\gtrsim 20$~nm, even low-intensity monochromatic EUV synchrotron radiation can induce multiple excitations in one droplet. Using the advanced techniques of high-resolution electron spectroscopy and photoelectron-photoion coincidence velocity-map imaging, the individual steps of the ICD process out of $^{1}$S and $^{3}$S states are unravelled at $h\nu$ below and up to a few eV's above the $E_i$. The main results obtained in this work can be summarized as follows:
i) At $h\nu$= 21.6 eV where the He nanodroplet is excited into the 1s2p-correlated absorption band, the highly resolved electron spectra and the perfectly isotropic distribution of the emitted electrons indicate that ICD takes place between two fully relaxed excited He atoms in metastable states that roam about the He droplet surface. Therefore, this type of ICD may be expected to be a slow process with a time constant on the order of 10 - 100~ps mainly determined by the roaming dynamics.
ii) The significant changes of the absorption spectrum of large He droplets point at nano-optical effects occurring in such large He nanodroplets (nano-focusing, resonance enhancement of the radiation). 
iii) ICD is efficient even at photon energies exceeding the adiabatic ionization energy of He droplets and above $E_i$ up to a few eV's due to electron-ion recombination into excited He states.
vi) The electron spectra measured in coincidence with He$^+$ and He$_2^+$ show that $^{1}$S ICD occurs by electronic relaxation in the entire range of resonant photoexcitation, even at $h\nu$ exceeding $E_i$ by about 1~eV. 
v) Droplet-induced electronic relaxation of excited He evolves into electron-ion recombination by which mainly He$^*$ atoms populated in triplet states are formed. 
vi) In the electron spectra recorded in coincidence with He$_2^+$, $^{3}$S ICD appears more prominently due to the enhanced ejection of He ions formed in this way; a crossing of the $^{3}$S ICD yield and the $^{1}$S ICD yield occurs when tuning $h\nu$ across the $E_i$. 
vii) The strongly differing abundances of He$^+$ and He$_2^+$ products for $^{1}$S \textit{vs.} $^{3}$S ICD indicate different scenarios of ICD taking place at the surface or in the bulk of the droplets, respectively. viii) ICD involving excimer He$_2^*$'s occurs only at $h\nu \geq E_i^\mathrm{drop}$ due to formation of stabilized He$_2^*$'s by electron-ion recombination.

\begin{figure}[!h]
	\centering
	\includegraphics[width=1.01\columnwidth]{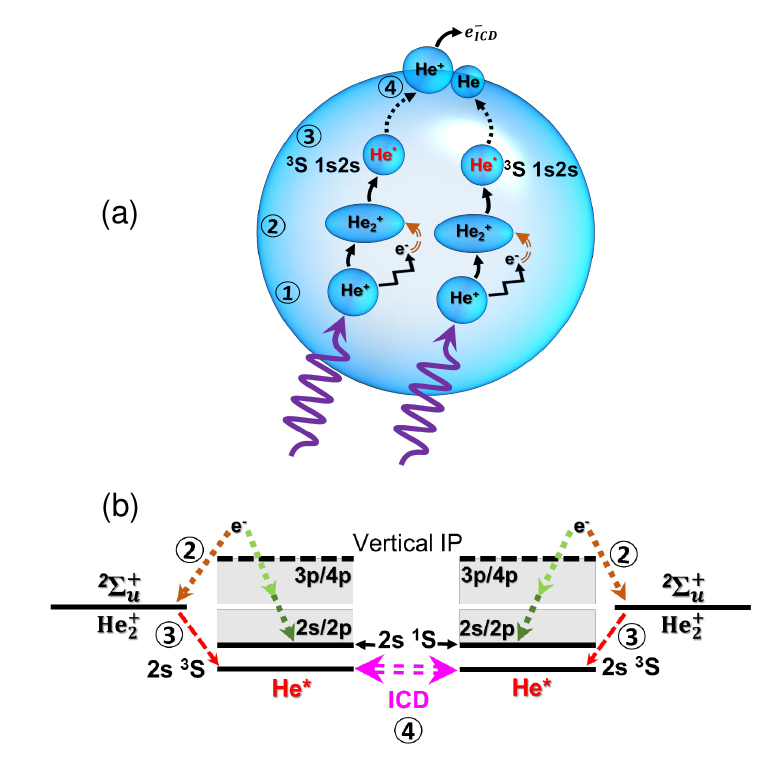}\caption{\label{Fig9} (a) Schematic illustration of ICD induced by electron-ion recombination in large He droplets following absorption of two ionizing EUV photons. (1) The emitted photoelectrons undergo elastic scattering inside the droplets and lose their kinetic energy. Depending on the excursion time of the electron, the He$^+$ photoions form stable He$_2^+$ dimers before recapturing the decelerated electrons (2). Once the electrons have recombined with the ions, excited He$_2^*$ dimers form which dissociate into one neutral He atom and one excited He atom in the $^3$S state (3). The two $^3$S He$^*$ atoms tend to migrate to the surface of the droplet where they decay by ICD (4). (b) Energy level diagram illustrating the dynamics (2), (3) and (4) leading to ICD as shown in (a).}
\end{figure}

The individual steps of the $^{3}$S ICD process which is occurring at photon energies exceeding $E_i$ due to electron-ion recombination are schematically illustrated in Fig.~\ref{Fig9}. Following photoionization of two He atoms, the emitted electrons perform a diffusion-like motion inside the droplets by which they loose their kinetic energy. Owing to the long-range Coulomb attraction, the electrons are drawn back to the ions which tend to form He$_2^+$ dimer ions by interaction with the surrounding He. Electron-ion recombination then leads to the formation of $^3$S-excited He$^*$ atoms or $^3\Sigma$-excited He$_2^*$ excimers. These metastable species tend to be expelled toward the droplet surface where they meet and decay by ICD.

To assess the general relevance of this process, other types of nanosystems should be studied in a similar size range, \textit{e.~g.} heavier rare-gas clusters and molecular clusters such as water clusters and nanodroplets. Some of these can be resonantly excited and ionized with conventional lasers~\cite{Serdobintsev:2018}. Electron-ion recombination to create highly reactive excited species may play an even more important role in bulk liquids and biological systems when exposed to ionizing radiation.

\section*{Acknowledgement}
M.M. and L.B.L. acknowledge financial support by Deutsche Forschungsgemeinschaft (project BE 6788/1-1), by the Danish Council for Independent Research Fund (DFF) via Grant No. 1026-00299B and by the Carlsberg Foundation. We thank the Danish Agency for Science, Technology, and Innovation for funding the instrument center DanScatt. SRK thanks Dept. of Science and Technology, Govt. of India, for support through the DST-DAAD scheme and Science and Eng. Research Board.  SRK acknowledges support for this research through the Indo-French Center for Promotion of Advanced Research (CEFIPRA). SRK, KS and SD acknowledge the support of the Scheme for Promotion of Academic Research Collaboration, Min. of Edu., Govt. of India, and the Institute of Excellence programme at IIT-Madras via the Quantum Center for Diamond and Emergent Materials. SRK gratefully acknowledges support of the Max Planck Society's Partner group programme, and M.M. and S.R.K. acknowledge funding from the SPARC Programme, MHRD, India. The research leading to this result has been supported by the project CALIPSOplus under grant agreement 730872 from the EU Framework Programme for Research and Innovation HORIZON 2020 and by the COST Action CA21101 ``Confined Molecular Systems: From a New Generation of Materials to the Stars (COSY)''.


%

 \end{document}


\title{Supplementary Materials\linebreak
Interatomic Coulombic decay induced by electron-ion recombination in large He nanodroplets}

\author{L. Ben Ltaief \textit{et al.}}

\maketitle









\date{}

\section{Dependency of the ICD electron yield on photon flux}
\begin{figure}[!h]
	\center
	\includegraphics[width=0.6\columnwidth]{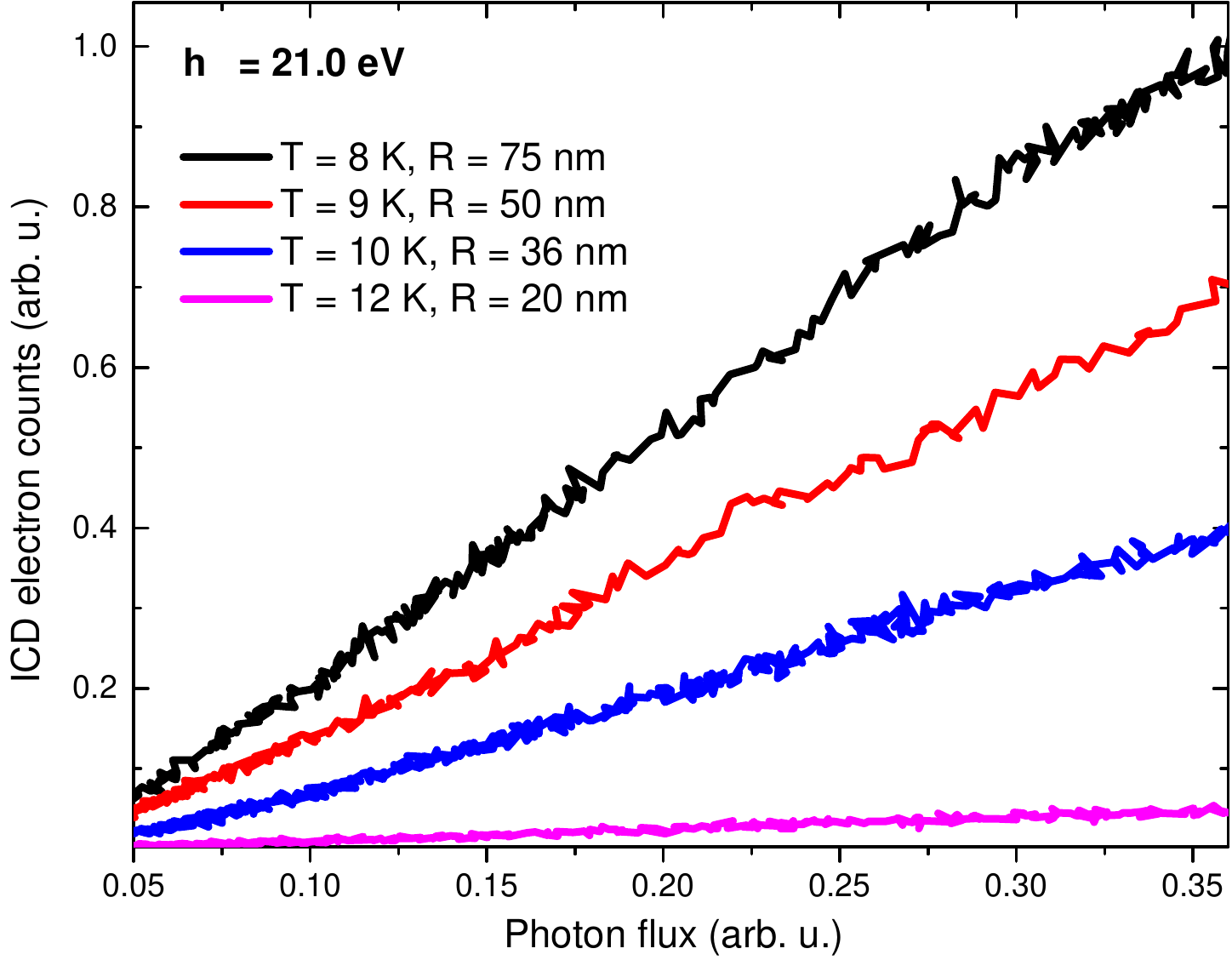}\caption{\label{SM_Fig1} Yield of ICD electrons as a function of photon flux measured at $h\nu = 21.0$~eV for He droplets of various droplet sizes set by controlling the temperature $T$ of the He nozzle. The photon flux was measured as photocurrents at a photodiode and at a mesh inserted into the photon beam.}
\end{figure}

To probe the dependency of the ICD electron yield on the intensity of the photon beam we recorded the total yield of ICD electrons with the HEA in the electron energy range $E_e = 16.2$ - 16.7~eV at $h\nu = 21.0$~eV and for different droplet sizes $R$, see SM Fig.~1. The photon flux was varied by gradually opening and closing the exit slit of the monochromator of the beamline. It was measured using a photodiode placed at the end of the beamline as well as by measuring the current at a mesh placed into the photon beam. The latter two currents were perfectly proportional to one another. In the range of lowest photon flux where the slit was nearly fully closed the ICD electron yield appears to show a slightly nonlinear dependency on the photon flux. This may be due to changes in the size and shape of the intensity distribution in interaction region given by diffraction effects of the light passing through the narrow slit.

\section{Ion kinetic energy distributions}
\begin{figure}[!h]
	\center
	\includegraphics[width=0.8\columnwidth]{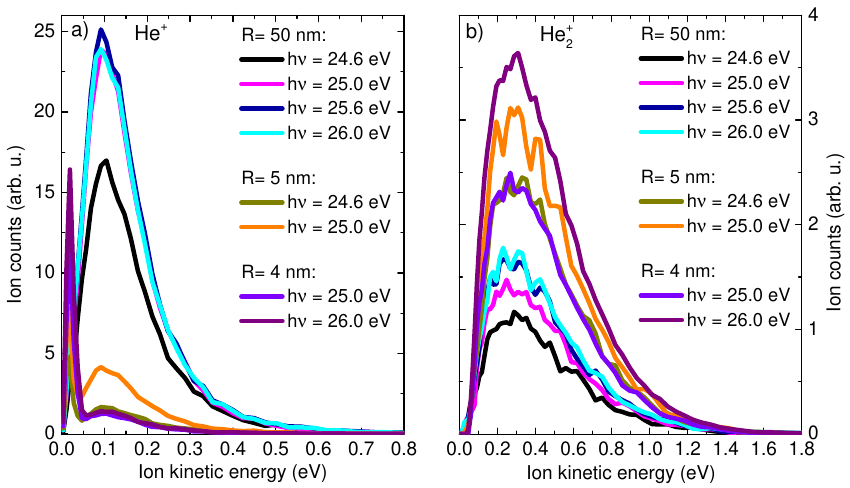}\caption{\label{SM_Fig2} Ion kinetic energy distributions of He$^+$ [panel a)] and He$_2^+$ [panel b)] measured for different droplet conditions and different photon energies $h\nu$ at and above $E_i=24.6$~eV. All the ion spectra in a) and b) are background subtracted and normalized to the EUV photon flux.}
\end{figure}

A compilation of He$^+$ and He$_2^+$ ion kinetic energy distributions measured for various He nanodroplet radii $R$ and photon energies $h\nu$ is shown in SM Fig.~2. The He$_2^+$ spectra have the same shape up to variable amplitude; they all feature a maximum around 0.3~eV and a broad tail that extends up to 1.6~eV. This generic kinetic energy distribution of He$_2^+$ was previously observed and interpreted by an impulsive ejection mechanism~\cite{callicoatt:1998}. The only remarkable trend is a dropping amplitude for large He droplets ($R=50$~nm). This is likely due to the tendency of large He droplets to efficiently trap ions as any ion tends to solvate in liquid He by forming a dense solvation complex~\cite{asmussen2023dopant}. 

The He$^+$ spectra feature two peaks: A sharp one near zero kinetic energy, which is most prominent for small droplets ($R=4$~nm), and a broad one peaked around 0.1~eV, which is present for all values of $R$ and $h\nu$. The former is due to free He atoms accompanying the He droplet jet, whereas the latter is characteristic of photoionization of He nanodroplets. For small nanodroplets the two-photon ionization probability is negligibly small. Therefore, Coulomb explosion of two He$^+$ photoions created in the same droplet can be ruled out. Moreover, one would expect a shift of the most probable energy of ejected He$^+$ ions as a function of droplet size as ions created in the bulk of large droplets would likely undergo binary collisions leading to a reduction of their kinetic energy~\cite{Shcherbinin:2017}. However, the 0.1-eV feature remains unchanged up to amplitude variations. 

Therefore, we rationalize the observation of He$^+$ ions with kinetic energies peaked at 0.1~eV by photoionization of nearest-neighbor pairs of He atoms into the repulsive $A$ state  the $He_2^+$ molecular ion, see the blue line in Fig.~6 in the main text. The transition from the $He_2$ ground state to the $A$ state is forbidden in the free He$_2$ system. However, it may become partly allowed due to symmetry breaking when the transition takes place in the He droplet environment. The estimated ion kinetic energy released following the dissociation along the $A$-state potential curve is 0.06~eV, which is in decent agreement with the experimental value (0.1~eV).

\section{Electron spectra}
\begin{figure}[!h]
	\center
	\includegraphics[width=0.85\columnwidth]{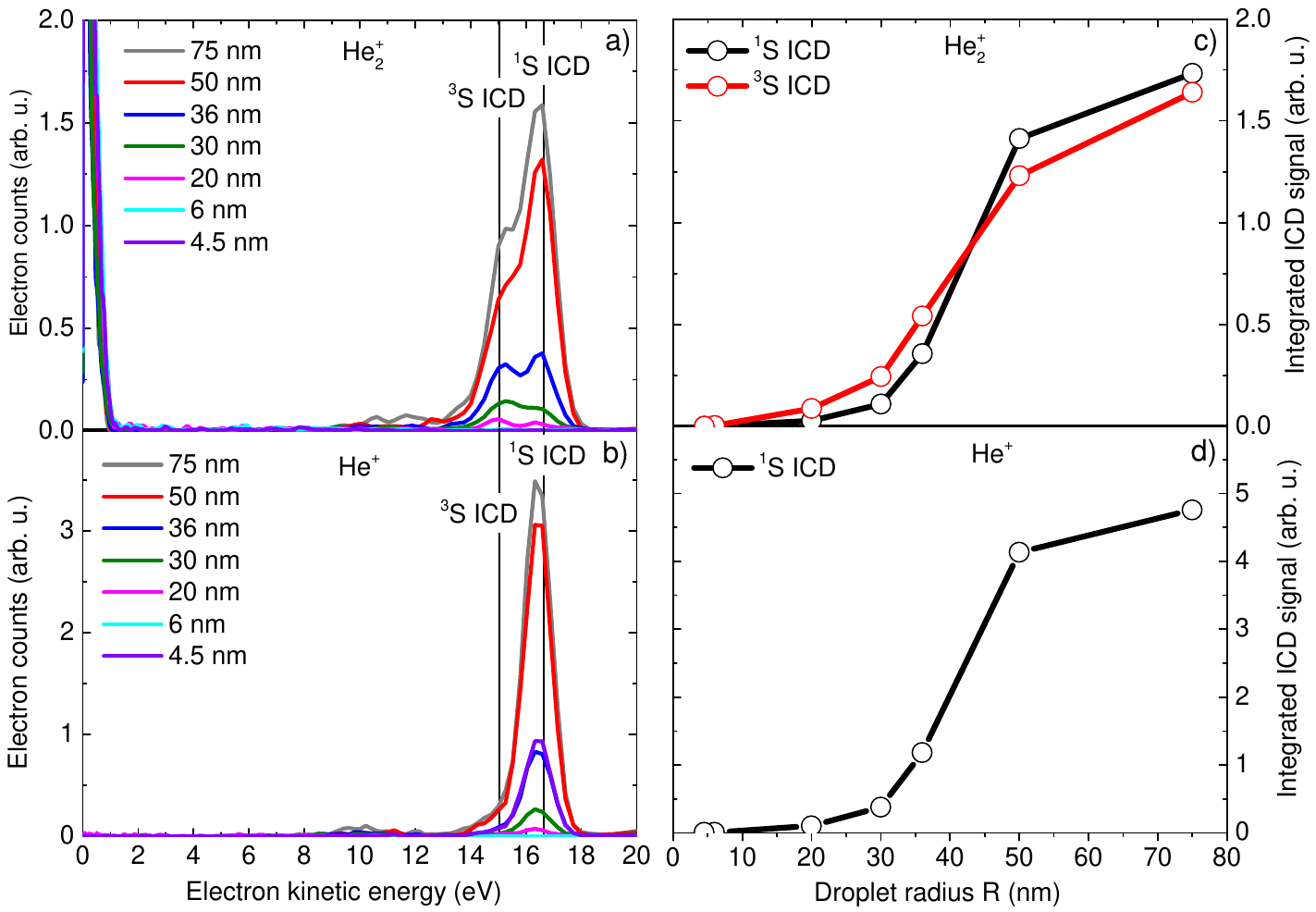}\caption{\label{SM_Fig3} a) \& b) Electron spectra of He nanodroplets measured in coincidence with He$_2^+$ and He$^+$, respectively, for different He droplet sizes and at $h\nu = 23.8$ eV. c) \& d) Integrated $^1$S and $^3$S ICD signals measured in coincidence with He$_2^+$ and He$^+$, respectively, as a function of droplet radius.}
\end{figure}

In addition to Fig.~5 in the main text, further information on the droplet size-dependent ICD electron yield for pairs of $^1$S and $^3$S He atoms can be obtained from the electron spectra shown in SM Fig.~3. Panels a) and b) show electron spectra of He nanodroplets of variable droplet sizes ($R=4.5$~nm up to 75~nm) measured at $h\nu = 23.8$~eV in coincidence with He$_2^+$ and He$^+$, respectively. Clearly, ICD starts to occur for droplets with radius $R\gtrsim 20$~nm and becomes more and more pronounced when the He droplet size increases. Electron spectra measured in coincidence with He$^+$ exhibit only one feature at 16.6~eV which is assigned to ICD out of the $^1$S state, whereas the electron spectra measured in coincidence with He$_2^+$ show two features; one $^1$S ICD feature at 16.6~eV and one at 15.0~eV attributed to ICD out of the $^3$S state. The droplet-size dependence of the integrated $^1$S and $^3$S ICD electron yields are shown in SM Fig.~3 c) and d). Surprisingly, the $^3$S ICD feature in the He$_2^+$ coincidence spectra appears more pronounced at smaller droplet sizes whereas at larger droplet sizes $R\gtrsim 40$~nm, the $^1$S ICD feature again dominates. 

\section{Relative ICD intensity}
The relative experimental ICD intensities $I_{ICD}[\mathrm{He}^+]$ and $I_{ICD}[\mathrm{He}_2^+]$ plotted in Fig.~7~b) in the main text are obtained from the calculated ratios between the integrated ICD electron yields $S$ measured in coincidence with He$^+$ and He$_2^+$, respectively, and of the photoline. Both of these ratios contain correction factors to account for the second-order radiation ($2h\nu$) present it the photon beam,
 \begin{equation}
    I_{ICD}[\mathrm{He}^+]=\frac{S_{ICD}[\mathrm{He}^+]}{S_{pl}[\mathrm{He}^+]} - \epsilon_{1}\frac{S_{pl (2h\nu)}[\mathrm{He}^+]}{S_{pl}[\mathrm{He}^+]},
\end{equation}
 \begin{equation}
    I_{ICD}[\mathrm{He}_2^+]=\frac{S_{ICD}[\mathrm{He}_2^+]}{S_{pl}[\mathrm{He}_2^+]} - \epsilon_{2}\frac{S_{pl (2h\nu)}[\mathrm{He}_2^+]}{S_{pl}[\mathrm{He}_2^+]}.
\end{equation}
Here, $\epsilon_{1} = S_{ICD(h\nu^{'})}[\mathrm{He}^+]/S_{pl(h\nu^{'})}[\mathrm{He}^+]$ and $\epsilon_{2} = S_{ICD(h\nu^{'})}[\mathrm{He}_2^+]/S_{pl(h\nu^{'})}[\mathrm{He}_2^+]$ are taken from \cite{ltaief2023efficient} and defined as the efficiency of $ICD[\mathrm{He}^+]$ and $ICD[\mathrm{He}_2^+]$, respectively, at higher photon energy $h\nu^{'}=2h\nu\geq 44.4$~eV.

The relative experimental $^1$S and $^3$S ICD yields plotted in Fig.~7~c) as well as those plotted in Fig.~4~c) and d) are evaluated by
\begin{equation}
    I_{^1S-ICD} = \alpha_{^1S} \times I_{^1 S ICD}[\mathrm{He}^+] + \beta_{^1S} \times I_{^1 S ICD}[\mathrm{He}_2^+],
\end{equation}
\begin{equation}
    I_{^3S-ICD} = \alpha_{^3S} \times I_{^3S-ICD}[\mathrm{He}^+] + \beta_{^3S} \times I_{^3S-ICD}[\mathrm{He}_2^+].
\end{equation}
Here, $I_{^1S-ICD}[\mathrm{He}^+]$ and $I_{^3S-ICD}[\mathrm{He}^+]$ are obtained separately but in a similar way as in equation (1). $I_{^1S-ICD}[\mathrm{He}_2^+]$ and $I_{^3S-ICD}[\mathrm{He}_2^+]$ are obtained separately but in a similar way as in equation (2). $\alpha_{^1S}$, $\beta_{^1S}$, $\alpha_{^3S}$ and $\beta_{^3S}$ are weighting factors 
\begin{equation*}
    \alpha_{^1S} =  \frac{A_{^1S-ICD}[\mathrm{He}^+]}{A_{^1S-ICD}[\mathrm{He}^+] + A_{^1S-ICD}[\mathrm{He}_2^+]},
\end{equation*}
\begin{equation*}
 \beta_{^1 S} =  \frac{A_{^1S-ICD}[\mathrm{He}_2^+]}{A_{^1S-ICD}[\mathrm{He}_2^+] + A_{^1S-ICD}[\mathrm{He}^+]},
 \end{equation*}
\begin{equation*}
    \alpha_{^3 S} =  \frac{A_{^3S-ICD}[\mathrm{He}^+]}{A_{^3S-ICD}[\mathrm{He}^+] + A_{^3S-ICD}[\mathrm{He}_2^+]},
\end{equation*}
\begin{equation*}
 \beta_{^3 S} =  \frac{A_{^3S-ICD}[\mathrm{He}_2^+]}{A_{^3S-ICD}[\mathrm{He}_2^+] + A_{^3S-ICD}[\mathrm{He}^+]}.
 \end{equation*}
Here $A_{^1S-ICD}[\mathrm{He}^+]$ and $A_{^1S-ICD}[\mathrm{He}_2^+]$ denote the peak heights of the $^1$S ICD signals measured in coincidence with $\mathrm{He}^+$ and $\mathrm{He}_2^+$, respectively. $A_{^3S-ICD}[\mathrm{He}^+]$ and $A_{^3S-ICD}[\mathrm{He}_2^+]$ denote the peak heights of the $^3$S ICD signals measured in coincidence with $\mathrm{He}^+$ and $\mathrm{He}_2^+$, respectively.

%